\documentclass[preprint]{aastex63}
\usepackage{color}
\usepackage{bm}
\usepackage{cases}
\usepackage{multirow}
\usepackage{lineno}
\shorttitle{}
\shortauthors{}

\begin{document}

\title{Observations favor the redshift-evolutionary $L_X$-$L_{UV}$ relation of quasars from copula}

\author[0000-0003-3635-5375]{Bao Wang}
\affiliation{Department of Physics and Synergetic Innovation Center for Quantum Effects and Applications, Hunan Normal University, Changsha, Hunan 410081, China}
\affiliation{Purple Mountain Observatory, Chinese Academy of Sciences, Nanjing 210023, China}
\affiliation{School of Astronomy and Space Sciences, University of Science and Technology of China, Hefei 230026, China}

\author[0000-0003-2721-2559]{Yang Liu}
\affiliation{Department of Physics and Synergetic Innovation Center for Quantum Effects and Applications, Hunan Normal University, Changsha, Hunan 410081, China}

\author{Hongwei Yu}
\affiliation{Department of Physics and Synergetic Innovation Center for Quantum Effects and Applications, Hunan Normal University, Changsha, Hunan 410081, China}
\affiliation{Institute of Interdisciplinary Studies, Hunan Normal University, Changsha, Hunan 410081, China}

\author{Puxun Wu}
\affiliation{Department of Physics and Synergetic Innovation Center for Quantum Effects and Applications, Hunan Normal University, Changsha, Hunan 410081, China}
\affiliation{Institute of Interdisciplinary Studies, Hunan Normal University, Changsha, Hunan 410081, China}

%\correspondingauthor{Hongwei Yu}
\email{baowang@pmo.ac.cn}
\email{yangl@hunnu.edu.cn}
\email{hwyu@hunnu.edu.cn}
%\correspondingauthor{Puxun Wu}
\email{pxwu@hunnu.edu.cn}

%\linenumbers
\begin{abstract}

 We compare,  with  data from  the quasars, the Hubble parameter measurements, and the Pantheon+ type Ia supernova,  three different relations between X-ray luminosity ($L_X$) and ultraviolet luminosity ($L_{UV}$) of  quasars. These three relations consist of  the standard  and  two redshift-evolutionary $L_X$-$L_{UV}$ relations which are  constructed  respectively by considering a redshift dependent correction to the luminosities of  quasars and using the statistical tool called copula.  By employing  the PAge approximation for a cosmological-model-independent description of the cosmic background evolution and dividing the quasar data into the low-redshift and high-redshift parts,   we find that the constraints on the PAge parameters from the low-redshift and high-redshift data, which are obtained with the redshift-evolutionary relations,  are consistent with each other, while they are not when the standard relation is considered. If the data are used to constrain the coefficients of the relations and the PAge parameters simultaneously, then the observations support the redshift-evolutionary relations at more than $3\sigma$.   The Akaike and Bayes information criteria indicate that there is  strong evidence against  the standard relation and mild evidence against  the redshift-evolutionary relation constructed by considering a redshift dependent correction to the luminosities of  quasars. This suggests that the redshift-evolutionary $L_X$-$L_{UV}$  relation of quasars constructed from copula is favored by  the observations.

~\\

\end{abstract}

\section{Introduction}

Quasars are extremely luminous and persistent energy sources powered by  supermassive black holes. The  luminosities of quasars are so great that  the maximum redshift of quasars  can reach $z>7$~\citep{Mortlock2011, Eduardo2018, Lyke2020, Wang2021}. If quasars can be regarded as  standard candles, they  will cover the redshift desert of  cosmological  data, and may play an important role in understanding the property of dark energy and the possible origin of the Hubble constant ($H_0$) tension.  To use data from quasars for cosmological purposes, one needs to construct a luminosity relation to determine the distance of quasars.  Several empirical relations  have been proposed~\citep{Baldwin1977, Paragi1999, Chen2003, Watson2011, La2014, Wang2014, Cao2021}.
Among them,   the nonlinear relation between the X-ray luminosity ($L_X$) and the ultraviolet (UV) luminosity ($L_{UV}$) \citep{Risaliti2015, Risaliti2019, Lusso2016, Lusso2020} is a very popular one. The $L_X$-$L_{UV}$ relation has  been utilized to construct the Hubble diagram of the quasars up to $z\sim 7.5$, and  has been applied in the quasar cosmology widely \citep{Khadka2020a, Khadka2020b, Wei2020, Khadka2021, Li2021, Lian2021, Bargiacchi2022}.

With the $L_X$-$L_{UV}$ relation and the logarithmic polynomial expansion of the cosmic distance, it has been found that the distance modulus/redshift relation of the quasars at $z > 1.4$ has  a more than 4$\sigma$ deviation from the prediction of  the cosmological constant plus the cold dark matter ($\Lambda$CDM) model with $\Omega_{m0}=0.3$~\citep{Risaliti2019, Lusso2019} \footnote{It should be pointed out that controversies have arisen  regarding this deviation~\citep{Yang2020, Banerjee2021, Bargiacchi2021, Khadka2022, Sacchi2022, Khadka2022b, Dainotti2022, Lenart2023, Wang2022, Mehrabi2020, Velten2020, Petrosian2022}.
Some researches  indicated that this deviation may originate from the divergence of  the logarithmic  polynomial  expansion at the high-redshift regions~\citep{Yang2020, Banerjee2021}.}, where $\Omega_{m0}$ is the present dimensionless matter density parameter.
This deviation may indicate that   
 the standard $L_X$-$L_{UV}$ relation of quasars may  not be accurate.  
 Recently,   \citet{Khadka2022} found that part of the quasar data shows  evidence of redshift evolution of the $L_X$-$L_{UV}$ relation, suggesting a  possible redshift-evolutionary $L_X$-$L_{UV}$ relation.  \citet{Dainotti2022} first obtained a three-dimensional and redshift-evolutionary  $L_X$-$L_{UV}$ relation 
by considering a redshift dependent  correction to the luminosities of quasars.  We also constructed a three-dimensional and redshift-evolutionary  $L_X$-$L_{UV}$ relation by using the powerful statistical tool called copula~\citep{Wang2022}. For the standard and two redshift-evolutionary $L_X$-$L_{UV}$  relations, it remains interesting to determine which one is favored  by observations.

To compare three different $L_X$-$L_{UV}$  relations with observations, the theoretical value of the luminosity distance $d_L$ must be given.  One way to  achieve this goal is  to use a cosmological model  to derive the luminosity distance.  The obtained results will then be model-dependent. The other one is to resort to 
the  so-called cosmography, which is  cosmological-model-independent.  However, 
 the usual cosmographic method, which is based on  the Taylor expansion or the Pad\'{e} approximation of the Hubble expansion rate, suffers  the divergence problem in the high-redshift regions.
Recently, \citet{Huang2020} proposed the PAge approximation  and found  that it can describe the global expansion history of our universe with high accuracy.
The PAge approximation can be considered as a cosmological-model-independent method for the high-redshift cosmography \citep{Huang2020, Huang2021, Cai2022} and has been applied in cosmology widely.  For example, 
 the PAge approximation has been used to explore the supernova magnitude evolution \citep{Huang2020}, test the high redshift gamma-ray burst luminosity correlations \citep{Huang2021}, reaffirm the cosmic expansion history \citep{Luo2020}, and investigate the $S_8$ tension \citep{Huang2022} and the $H_0$ tension \citep{Cai2022a, Cai2022b}. 
More recently, by dividing the quasar data into the low-redshift and high-redshift parts and using them to constrain the coefficients of the $L_X$-$L_{UV}$ relation and the PAge parameters, \citet{Li2022} found that there are apparent inconsistencies between the results from the low-redshift data and the high-redshift ones respectively, and concluded that the deviation between the distance modulus/redshift relation of the high-redshift quasars and the prediction from the $\Lambda$CDM model found in Ref. \citep{Risaliti2019} probably originates from the redshift-evolutionary  effects of the  $L_X$-$L_{UV}$ relation.  

In this paper, we check whether the $L_X$-$L_{UV}$ relation is redshift-evolutionary  and compare three different $L_X$-$L_{UV}$ relations \citep{Dainotti2022, Wang2022, Risaliti2015}  by using the PAge approximation with the observational data comprising of  the 2421 quasar sample~\citep{Lusso2020}, the Hubble parameter ($H(z)$) measurements~\citep{Moresco2020} and the Pantheon+ type Ia supernova (SNe Ia) data~\citep{Brout2022}. We find that  the three-dimensional and redshift-evolutionary $L_X$-$L_{UV}$  relation from copula~\citep{Wang2022} is favored by the observational data. 

The rest of the paper is organized as follows.
In \autoref{sec:2}, we introduce the  PAge approximation,  the three different $L_X$-$L_{UV}$ relations and the data sets.
The redshift-evolutionary effects of $L_X$-$L_{UV}$ relations are examined in \autoref{sec:3}.
A comparison of  the three different relations is given in \autoref{sec:4} and 
the selection effects are discussed in \autoref{sec:5}. We summarize  our conclusions  in \autoref{sec:6}.

%-------------------------------------------

\section{Models and Data}\label{sec:2}
\subsection{PAge Approximation}

The PAge approximation~\citep{Huang2020} is a cosmographic method proposed to describe the cosmic expansion history.  The detailed process to construct the PAge approximation is given  in the Appendix~(\ref{Appendix:PAge}).
Compared with the Taylor expansion and the Pad\'{e} approximation, the PAge approximation is accurate enough to perform well in the high-redshift cosmography~\citep{Huang2021, Cai2022}.
In the PAge approximation, the  Hubble expansion rate $H$ is taken as a function of the cosmic time $t$:
\begin{equation}\label{eq:page-approximation}
	\frac{H(t)}{H_{0}}=1+\frac{2}{3}\left(1-\eta \frac{H_{0} t}{p_{\rm age}}\right)
	\left(\frac{1}{H_{0} t}-\frac{1}{p_{\rm age}}\right),
\end{equation}
where  parameters $p_{\rm age}$ and $\eta$ are defined, respectively,  as
\begin{equation}\label{eq:page}
	p_{\rm age}=H_0 t_0,
\end{equation}
and
\begin{equation}\label{eq:eta}
	\eta=1-\frac{3}{2} p_{\rm age}^2 (1+q_0).
\end{equation}
Here $t_0$ is the cosmic age and $q_0$ is the present deceleration parameter.
According to the definition of $H(t)$: $H(t)=\frac{1}{a(t)}\frac{{\rm d}a(t)}{ {\rm d} t}$ with $a(t)$ being the cosmic scale factor, Eq.~(\ref{eq:page-approximation}) can be rewritten  as a differential equation.
Solving this differential equation, one can obtain the relation between  $a(t)$ and $t$: 
\begin{equation}\label{Scale factor}
	a(t)=\left( \frac{t}{t_0} \right)^{\frac{2}{3}}
	\exp \left[\frac{1}{3 t_0^2}(t-t_0) \left(\eta  t + 3 H_0 t_0^2-\eta t_0 -2 t_0\right)\right].
\end{equation}
Apparently, the expansion of the universe can be described by three parameters: $t_0$, $H_0$ and $q_0$.
Thus, the PAge approximation provides  a cosmological-model-independent method to describe   the cosmic background evolution.
With the PAge approximation, the luminosity distance $d_L$ has the form
\begin{equation}\label{dL}
		d_L(z) = (1+z) \int^{t_0}_{t_z} {\frac{{\rm d}t}{a(t)}} ,
\end{equation}
where $t_z$ is the time of light emission, and  $z=\frac{1}{a(t)}-1$ is the redshift. The relation between  $t_z$ and $z$ can be inferred from Eq.~(\ref{Scale factor}).

\subsection{X-ray and UV Luminosity Relations}\label{relation}

\citet{Risaliti2015} discovered that there is a nonlinear relation between the X-ray luminosity and the  UV luminosity of quasars, which takes the form
\begin{eqnarray}\label{relation1}	\log(L_X)=\beta + \gamma \log(L_{UV}),
\end{eqnarray}
where $\beta$ and $\gamma$ are two constants.

 Recently,  by    using the Gaussian copula, we have constructed a three-dimensional and redshift-evolutionary X-ray and UV luminosity  relation~\citep{Wang2022} 
\begin{equation}\label{3D}
	\log(L_X)=\beta + \gamma \log(L_{UV})+\alpha \ln(\bar{a}+z)
\end{equation}
with $\bar{a}=5$, where $\alpha$ is a new parameter that characterizes the redshift-evolution of the relation and $\ln \equiv \log_e$.   In deriving the above three-dimensional luminosity relation, we have assumed that $\log(L_{UV})$  and  $\log(L_X)$ satisfy, respectively,  the following Gaussian distributions 
\begin{eqnarray}\label{GP}
f(x)=\frac{1}{\sqrt{2\pi}\sigma_x} e^{-\frac{(x-\bar{a}_x)^2}{2\sigma_x^2}}, \quad f(y)=\frac{1}{\sqrt{2\pi}\sigma_y} e^{-\frac{(y-\bar{a}_y)^2}{2\sigma_y^2}}, 
\end{eqnarray}
where $x = \log(L_{UV})$, $y = \log(L_X)$,   $\bar{a}_x$ and  $\bar{a}_y$ represent the mean value, and $\sigma_x$ and $\sigma_y$ are the standard deviations.  Furthermore, the probability density distribution of the quasars needs to be known to construct the three-dimensional luminosity relation. \cite{Lusso2020} have pointed out that  the 2421 quasar data points satisfy the Gamma distribution in the $z$ space. Since the Gaussian distribution will give a simple expression of the X-ray and UV luminosity relation and  the log-transformation is a common way to transform the non-Gaussian distribution  into the Gaussian one, 
we consider the $z_*$ space,   where  $z_*\equiv \ln(\bar{a}+z)$ with $\bar{a}$ being a constant, 
and find that  
\begin{eqnarray}\label{fz}
f(z_*)=\frac{1}{\sqrt{2\pi}\sigma_{z_*}} e^{-\frac{(z_*-\bar{a})^2}{2\sigma_{z_*}^2}}
\end{eqnarray}
can describe approximately the redshift distribution of the quasars.  Utilizing  the Gaussian copula and  Eqs.~(\ref{GP}, \ref{fz}), \cite{Wang2022} constructed the relation shown in Eq.~(\ref{3D}).

 In Eq.~(\ref{3D}) , the standard relation is recovered when $\alpha=0$. If  $\bar{a}=1$, the relation given in Eq.~(\ref{3D}) reduces to the one proposed in~\citep{Dainotti2022}  by assuming that the luminosities of quasars are corrected by a redshift-dependent function $(1+z)^\alpha$.  In~\citep{Dainotti2022}, the value of $\alpha$ is determined by using the EP method~\citep{Efron1992}, while in our analysis, $\alpha$ is treated as a free parameter.
To avoid confusion, we name the relations with  $\bar{a}=5$  and  $\bar{a}=1$ as  Type I and Type II, respectively.
Converting  luminosity to  flux, we can define a  function $\Phi$ via
\begin{eqnarray}\label{flux}
	\log(F_X)&=& \Phi(\log(F_{UV}),d_L) \nonumber \\
			&=& \beta + \gamma \log(F_{UV}) +\alpha \ln(\bar{a}+z)+(\gamma-1) \log(4 \pi d_L^2)  .
\end{eqnarray}
Here $F_X= \frac{L_X}{4\pi d_L^2}$ and $F_{UV}=\frac{L_{UV}}{4\pi d_L^2 }$ are the fluxes of the X-ray  and UV respectively.

\subsection{Data Sets}\label{data}

To compare three different $L_X$-$L_{UV}$ relations, we will use the latest quasar data. Furthermore,  we also consider  the Hubble parameter measurements and Pantheon+ SNe Ia sample in order to  constrain the PAge parameters and the coefficients of the relations  tightly and break the degeneracy between different parameters.
The data sets are  as follows.

\begin{itemize}

	\item Quasars

	The data of quasars in our analysis comprise of 2421 X-ray and UV flux measurements~\citep{Lusso2020}, which cover a redshift range of $z \in [0.009, 7.541]$.
The relation coefficients and   the PAge parameters  can be obtained by maximizing  
$-\ln \mathcal{L}$, where $\mathcal{L}$ is the D'Agostinis likelihood function \citep{Agostini2005}
\begin{equation}\label{Lc}
	\mathcal{L}_Q(\delta,\beta,\gamma,\alpha, {\bf p})\propto\prod_{i} \frac{1}{\sqrt{2 \pi (\delta^2+\sigma_{{\rm tot}, i}^2 )}}		\exp\left \{-
\frac{[\log(F_X)_i-\Phi(\log(F_{UV}),d_L({\bf p}))_{i}]^2}{2\left(\delta^2+\sigma_{{\rm tot}, i}^2 \right) }  \right\}.
\end{equation}
Here $\sigma_{{\rm tot}, i}=\sigma_{X,i}^2 + \gamma^2\sigma_{UV,i}^2$ represents the total measurement error in $\log(F_X)$ and $\log(F_{UV})$, $\delta$ is  the intrinsic dispersion,  function  $\Phi$ is defined in Eq.~(\ref{flux}),  $d_L$ is the luminosity distance predicted by the PAge approximation in Eq.~(\ref{dL}) and $\bf p$ represent the PAge parameters.
	
	\item Hubble parameter measurements
	
	The updated 32 $H(z)$ measurements have a redshift range of $z \in [0.07,1.965]$~\citep{Moresco2020, Liu2023}, which contain 17 uncorrelated and 15 correlated measurements.
	The  15 correlated measurements are from Refs. \citep{Moresco2012, Moresco2015, Moresco2016} and  their covariance matrices are given in~\citep{Moresco2020}\footnote{All $H(z)$ measurements and their errors can be found at \href{https://gitlab.com/mmoresco/CCcovariance/}{https://gitlab.com/mmoresco/CCcovariance/}.}.
	The  $\chi^2$ of the $H(z)$ data is
	 \begin{equation}\label{likelihoodHz}
			\chi^2_H=\sum_{i=1}^{17}{\frac{[H_{{\rm th},i}(t)-H_{{\rm obs},i}(t)]^2}{\sigma_{H,i}^2}}+
			\Delta \hat{H}^{T} {\bf C^{-1}_{H}} \Delta \hat{H} ,
	\end{equation}
	where $H_{\rm obs}(t)$ is the observed value of the Hubble parameter, 
	$H_{\rm th}(t)$ represents the theoretical value from the PAge approximation, which is given in Eq.~(\ref{eq:page-approximation}), $\sigma_{H}$ is the observed uncertainty,  	$\Delta \hat{H}=H_{\rm th}(t)-H_{\rm obs}(t)$ is the data vector of the  15 correlated measurements,
	 and $\bf C^{-1}_{H}$ is the inverse of the covariance matrix.

	\item SNe Ia
	
	The updated Pantheon+ SNe Ia sample contains 1701  data points~\citep{Brout2022}.
	The nearby ($z < 0.01$) supernova data will not be used in our work since they are sensitive to the peculiar speed \citep{Brout2022}.
	 Thus, our analysis only includes 1590 data points in the redshift range of $z \in [0.01, 2.2613]$.
	The distance modulus $\mu_{\rm SN}$ of SNe Ia is defined as 
	\begin{equation}\label{mSN}
	\mu_{\rm SN}= m-M_B,
	\end{equation}
	where $m$ is the apparent magnitude and $M_{B}$ is the absolute magnitude.
The theoretical value of the distance modulus relates to 	 the luminosity distance $d_L$ through 
	\begin{equation}\label{mth}
	\mu_{\rm th}= 5\log \left ( \frac{d_{L}}{\mathrm{Mpc}}  \right ) +25.
	\end{equation}
	Then the $\chi^2_{\rm SN}$ of SNe Ia  has the form
	\begin{equation}
	\chi^2_{\rm SN}=\Delta \mu^T {\bf C^{~~-1}_{SN}} \Delta\mu,
	\end{equation}
	where $\Delta\mu=\mu_{{\rm SN}}-\mu_{{\rm th}}$ and $\bf C^{~~-1}_{SN}$ is the inverse of the covariance matrix which contains the systematic and statistical uncertainties.

\end{itemize}

The constraints on the relation coefficients,   the PAge parameters, the intrinsic dispersion and the absolute magnitude  from all observational data can be obtained by maximizing $\mathcal{L}= e^{-\frac{1}{2}(\chi^2_H+\chi^2_{\rm SN})} \mathcal{L}_Q$.

%-------------------------------------------
~\\

\section{Tests the redshift-evolutionary effects of relations}\label{sec:3}

After dividing all the quasar data into the high-redshift part and the low-redshift one, 
\citet{Li2022} used them to constrain the standard $L_X$-$L_{UV}$ relation with the PAge approximation and found that there exists tension between the results from the high-redshift and low-redshift data respectively. They concluded that this tension may  originate from the redshift-evolutionary effects of the $L_X$-$L_{UV}$ relation.
In this section, we discuss the constraints on the two redshift-evolutionary  $L_X$-$L_{UV}$ relations with the PAge approximation using data  from the low-redshift ($z<1.4$) and high-redshift ($z>1.4$) quasars respectively. 
 For  comparison,  the standard $L_X$-$L_{UV}$ relation is also considered in our analysis. 
Since the Hubble constant $H_0$  is degenerate with parameter $\beta$ in the relations and the quasars have large dispersions, we  also include  the $H(z)$ measurements and the SNe Ia data.

\begin{deluxetable}{ccccccccc}
	\caption{Marginalized constraints on the PAge parameters and the coefficients of the relations with 1$\sigma$ confidence level (CL) from the $H(z)$+SNe Ia+quasar data in the PAge approximation.  \label{Tab1}}
	\tablewidth{0pt}
	\tablehead{	 & \multicolumn{2}{c}{Type I  relation} & & \multicolumn{2}{c}{Type II relation} & & \multicolumn{2}{c}{Standard relation}\\
	\cline{2-3} \cline{5-6} \cline{8-9}
	& Low-redshift & High-redshift & & Low-redshift & High-redshift & & Low-redshift & High-redshift
	}
	\startdata
	$t_0$  & $13.84^{+0.50}_{-0.62}$ & $13.83^{+0.49}_{-0.55}$ & & $13.89^{+0.51}_{-0.65}$ & $13.91^{+0.47}_{-0.57}$ & & $13.76 \pm 0.52$ & $13.49^{+0.44}_{-0.54}$ \\
	$H_0$  & $69.2\pm 2.8$ & $69.1\pm 2.8$ & & $69.1\pm 2.8$ & $69.0\pm 2.4$ & & $68.2 \pm 2.6$ & $67.1^{+2.9}_{-2.5}$ \\
	$q_0$ & $-0.407 \pm 0.059$  & $-0.408^{+0.053}_{-0.060}$ & & $-0.404 \pm 0.056$ & $-0.398 \pm 0.059$ & & $-0.419 \pm0.059$ & $-0.488 \pm 0.053$ \\
	$M_B$ & $-19.357 \pm 0.086$ & $-19.362 \pm 0.0.89$ & & $-19.361 \pm 0.087$ & $-19.364 \pm 0.074$ & & $-19.389 \pm 0.084$ & $-19.432^{+0.098}_{-0.078} $ \\
	$\delta$ & $0.2357 \pm 0.0048$ & $0.2065\pm 0.0049$ & & $0.2357\pm 0.0048$ & $0.2058\pm 0.0054$ & & $0.2368\pm0.0048$ & $0.2124^{+0.0050}_{-0.0056}$ \\
	$\beta$ & $7.33^{+0.37}_{-0.41}$ & $8.42\pm 0.41$ & & $8.06\pm 0.45$& $9.59\pm 0.41$ & & $7.27 \pm 0.39$ & $8.24\pm0.40$ \\
	$\gamma$ & $0.606\pm 0.016$  & $0.551 \pm 0.015$ & & $0.604\pm 0.016$ & $0.548\pm 0.014$ & &  $0.632^{+0.014}_{-0.012}$ & $0.604\pm0.013$ \\
	$\alpha$ & $0.420 \pm 0.150$ & $0.739^{+0.096}_{-0.087}$ & & $0.131\pm0.047$ & $0.352\pm 0.045$ & & $-$ & $-$ \\
	\enddata
	\tablecomments{The unit of $t_0$ is Gyr and $H_0$ is ${\rm km\;s^{-1}\;Mpc^{-1}}$.}
\end{deluxetable}

\begin{figure}[h]
	\centering
	\includegraphics[width=.49\textwidth]{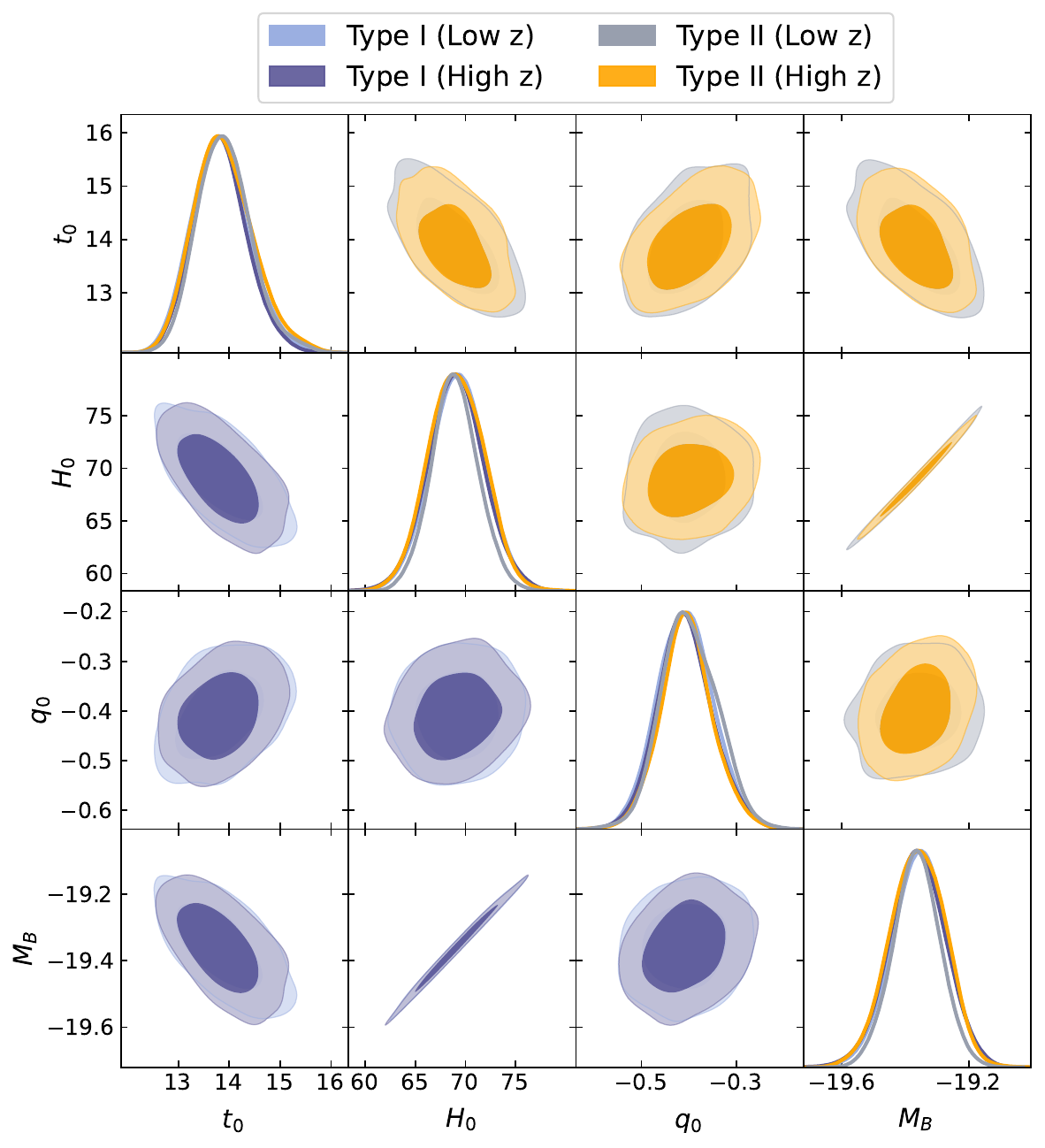}
	\includegraphics[width=.49\textwidth]{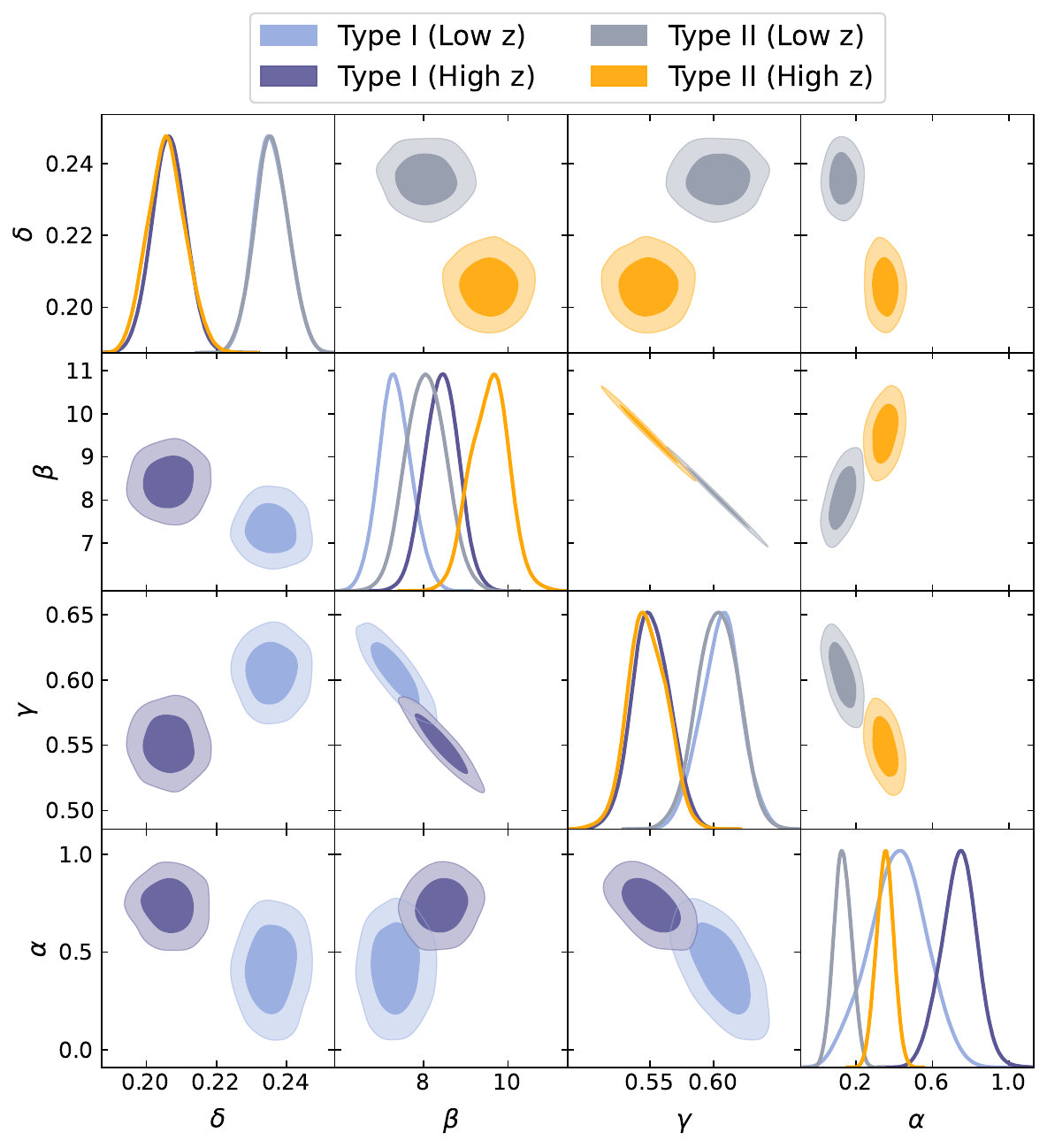}
	\caption{One-dimensional marginalized distributions and two-dimensional contours at 1$\sigma$ and 2$\sigma$ CLs  of the Type I and Type II  relations from low-redshift and high-redshift quasars.
		\label{fig1-1}
	}
\end{figure}

The best-fitting values with  uncertainties are obtained by using the Markov Chain Monte Carlo  (MCMC) method, which is available via the $emcee$ package in Python 3.7~\citep{Foreman2013}.
\autoref{Tab1} and \autoref{fig1-1} show the marginalized results and the probability contour plots, respectively.
From them,  we  obtain the following  main results:
\begin{itemize}
	\item For the standard relation, the constraints  from the low-redshift and high-redshift data  on all the parameters are  inconsistent. This agrees with the conclusion obtained in \citep{Li2022}.
	\item  For the redshift-evolutionary relations, the cosmographic parameters ($t_0$, $H_0$ and $q_0$) and the absolute magnitude ($M_B$) are consistent in two different redshift ranges. However,  the values of the relation coefficients ($\beta$, $\gamma$ and $\alpha$) and the intrinsic dispersion ($\delta$) from the low-redshift  and high-redshift data respectively are discrepant.  This is similar to what obtained in the standard relation case. 
	\item The redshift-evolutionary coefficient $\alpha$ deviates from zero at more than  $2 \sigma$, which
implies that the observational data support a redshift-dependent relation. Furthermore,  the redshift-evolutionary character is more apparent  in  the Type I  relation than that in the Type II case. 
\end{itemize}

Therefore,  the redshift-evolutionary  $L_X$-$L_{UV}$ relations are favored as opposed to the standard one since  consistent PAge parameters can be obtained in the redshift-evolutionary relations and the value of $\alpha$ deviates from zero at more than $2\sigma$.

%-------------------------------------------
~\\
\section{Simultaneous Constraints}\label{sec:4}

\begin{figure}
	\centering
	\includegraphics[width=1\textwidth]{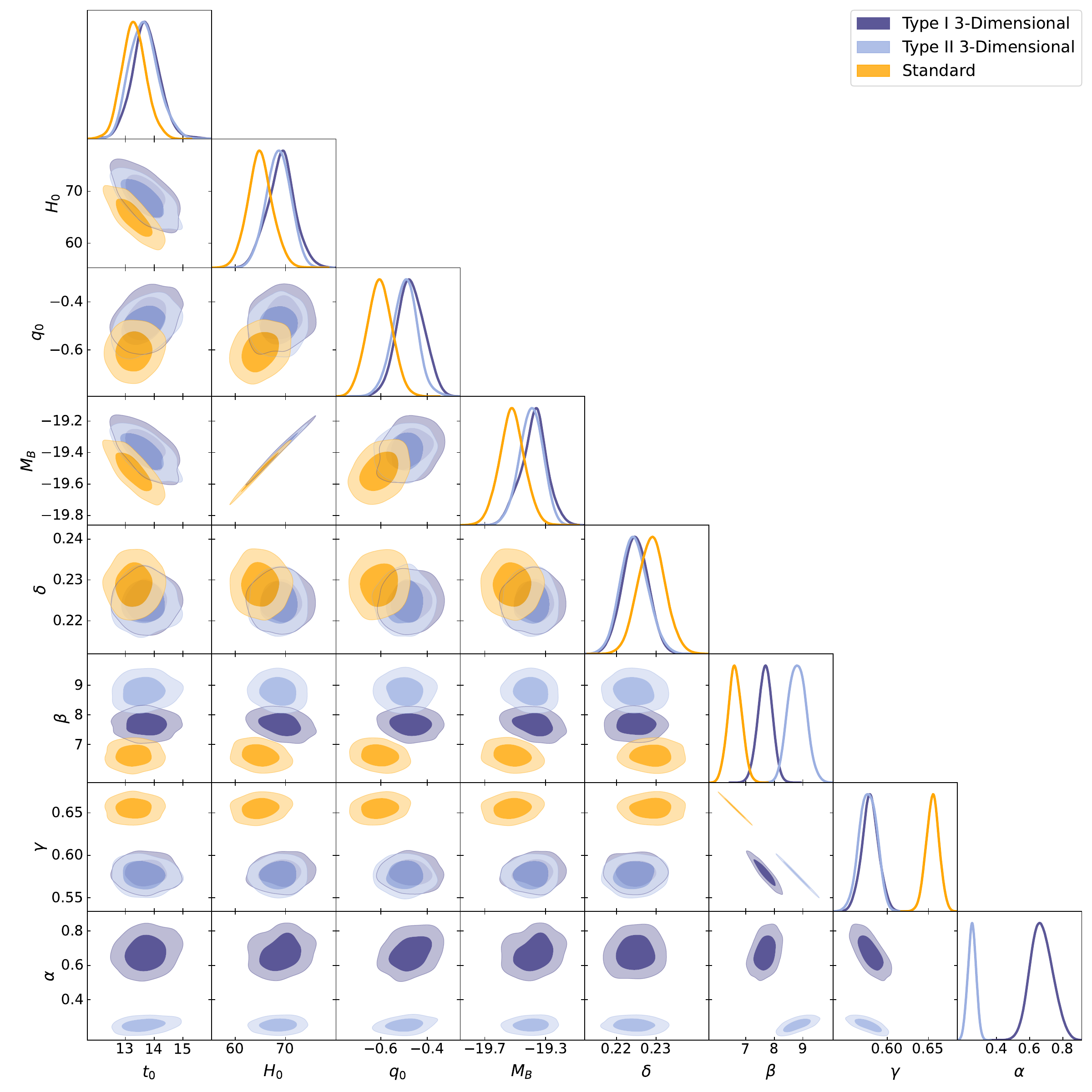}
	\caption{The marginalized posterior distributions and two-dimensional contours at 1$\sigma$ and 2$\sigma$ CLs from $H(z)$+SNe Ia+quasar data in the PAge approximation.
		\label{fig4-1}
	}
\end{figure}

To further compare the three different  $L_X$-$L_{UV}$ relations, we will use all $H(z)$, SNe Ia and quasar data to constrain the PAge parameters ($t_0$, $H_0$ and $q_0$), the absolute magnitude ($M_B$),  the coefficients of the relations ($\beta$, $\gamma$ and $\alpha$) and the intrinsic dispersion ($\delta$) simultaneously.
We will utilize the Akaike information criterion (AIC) \citep{Akaike1974, Akaike1981} and the Bayes information criterion (BIC) \citep{Schwarz1978} to find the relation favored by the observations. The AIC and BIC are defined, respectively,  as
\begin{eqnarray}\label{criterion}
\mathrm{AIC} & = & 2 p-2 \ln \mathcal{L}_{\rm max}\,, \\
\mathrm{BIC} & = & p \ln N-2 \ln \mathcal{L}_{\rm max}\, ,
\end{eqnarray}
where $\mathcal{L}_{\rm max}$ is the maximum value of the likelihood function, $p$ represents the number of free parameters and $N$ is the number of data. 
We calculate $\Delta$AIC(BIC) of the three relations, which denotes the difference in the AIC(BIC) of a given relation relative to the reference relation (the Type I relation ), i.e., $\rm \Delta AIC(BIC)=AIC(BIC)-AIC_{ref}(BIC_{ref})$, for the relation comparison.
If $0<\Delta \mathrm{AIC(BIC)}\le 2$, it is difficult to single out a better relation. If $2<\Delta \mathrm{AIC(BIC)}\le 6$ we have mild evidence against the given relation, and we have strong evidence against the given relation if $\Delta \mathrm{AIC(BIC)}>6$.
The parameter values and $\Delta$AIC(BIC) obtained from MCMC are summarized in \autoref{Tab2}, and the posterior distribution contours are shown in \autoref{fig4-1}.
 In the last column of \autoref{Tab2}, the constraints on the PAge parameters  and the absolute magnitude obtained from  data without the quasars are also displayed in order to assess the influence of the quasars.

\begin{deluxetable}{ccccccc}
	\tablecaption{The marginalized results with 1$\sigma$ CL for three different luminosity relations from $H(z)$+SNe Ia+quasar data in the PAge approximation.	 The results from data without the quasars are also displayed for a comparison.
	\label{Tab2}}
	\tablewidth{0pt}
	\tablehead{
		& Type I Three-Dimensional & Type II Three-Dimensional & Standard & Without quasars
	}
	\startdata
	$t_0$   & $13.72 \pm 0.50$ & $13.65^{+0.44}_{-0.56}$ & $13.30 \pm 0.43$ & $13.67^{+0.44}_{-0.53}$ \\
	$H_0$  & $69.1 \pm 2.8$ & $68.6 \pm 2.5$ & $65.0 \pm 2.5$ &  $69.4 \pm 2.7$ \\
	$q_0$ & $-0.471 \pm 0.059$  & $-0.496 \pm 0.055$ & $-0.606 \pm 0.054$ &  $-0.471 \pm 0.057$  \\
	$M_B$ & $-19.377^{+0.092}_{-0.079}$  & $-19.395 \pm 0.078$ & $-19.521 \pm 0.084$ &  $-19.369^{+0.091}_{-0.077}$ \\
	$\delta$ & $0.2248 \pm 0.0033$ & $0.2245^{+0.0032}_{-0.0037}$ & $0.2289\pm 0.0035$ & $-$ \\
	$\beta$ & $7.68\pm 0.25$ & $8.80\pm 0.32$ & $6.62\pm 0.24$ & $-$ \\
	$\gamma$ & $0.579\pm 0.010$  & $0.577 \pm 0.011$ & $0.655\pm 0.008$ & $-$ \\
	$\alpha$ & $0.672^{+0.063}_{-0.073}$ & $0.252 \pm 0.025$ & $-$ & $-$ \\
	\colrule
	$-2\ln \mathcal{L}_{\rm max}$ & $1246.550$ & $1250.415$ & $1350.201$ & $-$ \\
	$\Delta$AIC & $-$ & $3.865$ & $101.651$ & $-$ \\
	$\Delta$BIC & $-$ & $3.865$ & $95.346$ &\ $-$ \
	\enddata
	\tablecomments{The unit of $t_0$ is Gyr and $H_0$ is ${\rm km\;s^{-1}\;Mpc^{-1}}$.}
\end{deluxetable}

\begin{figure}[htp]
	\centering
	\includegraphics[width=.7\textwidth]{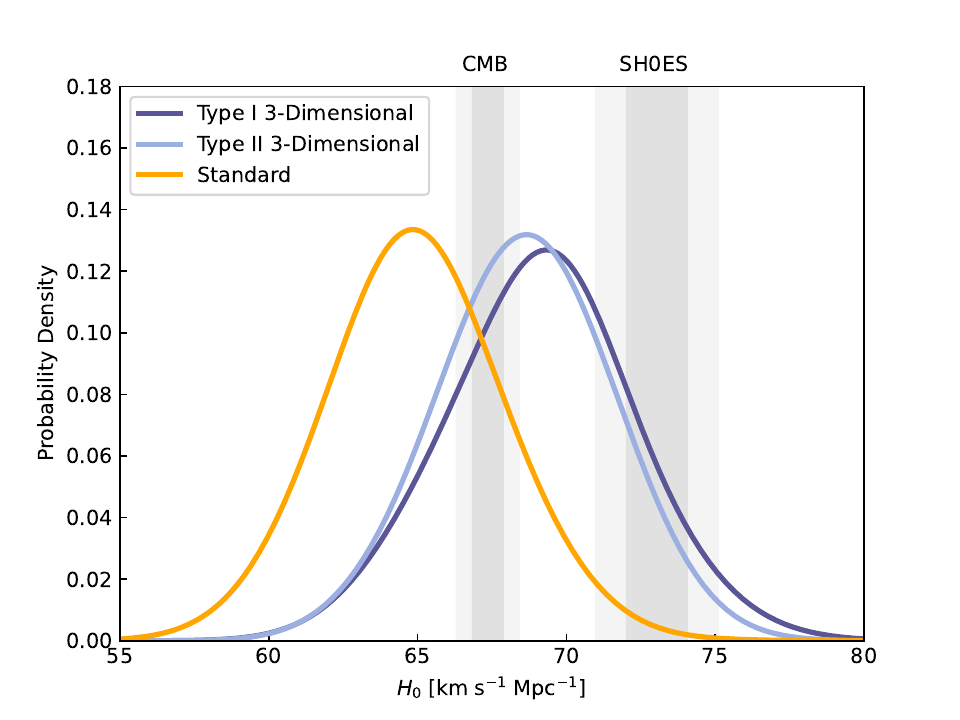}
	\caption{Probability density of Hubble constant $H_0$ for different luminosity relations.
	The gray shades represent the  results from the Planck 2018 CMB observations ($H_0=67.4\pm 0.5\; {\rm km\;s^{-1}\;Mpc^{-1}}$ \citep{Planck2020}) and  the SH0ES ($H_0=73.04\pm 1.04\; {\rm km\;s^{-1}\;Mpc^{-1}}$) \citep{Riess2022} at 68\% CL (dark gray) and 95\% CL (light gray). \label{fig4-2}
	}
\end{figure}

One can see that   the constraints on the PAge parameters ($H_0$, $t_0$ and $q_0$) and the absolute magnitude $M_B$ from  $H(z)$+SNe Ia+quasars with the Type I relation are well consistent with those from    data  without the quasars. The values of the PAge parameters  and the absolute magnitude  are  compatible   with each other for the Type I and Type II relations, while they apparently deviate for the standard relation.
The values of $t_0$ for the  Type I and Type II relations are more comparable with the result from the Planck 2018 CMB observations ($t_0=13.801\pm 0.024$ Gyr)~\citep{Planck2020} than those for the standard relation. The deceleration parameter $q_0$ for  the Type I and Type II relations is slightly larger than that in the $\Lambda$CDM model ($q_0=3\Omega_{m0}/2-1=-0.528\pm 0.011$) from the Planck 2018 CMB observations~\citep{Planck2020}, while its value for the standard relation is smaller than $q_0=-0.528\pm 0.011$. Furthermore, we find that the value of the Hubble constant $H_0$ for the Type I and Type II relations is lager than that  for the standard relation, and locates in the region  between the Planck 2018 CMB observations \citep{Planck2020} and the local distance ladders from SH0ES \citep{Riess2022}, which can be seen in \autoref{fig4-2}, where the values of $H_0$ from the quasars+$H(z)$+SNe Ia, the Planck 2018, and the SH0ES are shown.

The intrinsic dispersion $\delta$ and the relation coefficient $\gamma$ have almost the same values in the Type I and Type II relations,  which are larger than those obtained in the standard relation. The  value of $\gamma$ in the two three-dimensional relations is consistent with  $\gamma=0.586\pm 0.061$ obtained through the narrow redshift bin method~\citep{Lusso2020}. For the relation coefficient $\beta$, the Type II relation has the maximum value and the standard relation has the minimum one. The deviations of $\beta$ among three different relations are of more than $2\sigma$. The observations favor strongly the redshift-evolutionary relations since  $\alpha=0$ is ruled out at more than  $3\sigma$.     The value of $\alpha$ in the Type I relation is apparently larger than that in the Type II relation, which indicates that the redshift-evolutionary character in the Type I relation is more  apparent than that in the Type II relation.   
The $\alpha$ value in the Type I relation is comparable with   $\alpha=0.580^{+0.084}_{-0.099}$ obtained in  the $\Lambda$CDM model~\citep{Wang2022}.  From \autoref{Tab2}, we find that the Type I three-dimensional relation has the minimum AIC(BIC). The $\Delta$AIC(BIC) of the standard relation is much larger than 6, which indicates  that we have strong evidence against it. The $\Delta$AIC(BIC) of the Type II three-dimensional relation is 3.865. Thus, the  Type I three-dimensional relation has a slight advantage.

\vspace{\baselineskip}
%-------------------------------------------
\section{Selection effects}\label{sec:5}

To check whether  our result that the redshift-evolutionary  relation is favored by the observations is caused by the selection effects involved in the measurement process, we  follow the method given in \citep{Li2007,Singh2022} and  use firstly Monte Carlo simulations to simulate a realistic population of quasars as a function of redshift.    Since the 2421 quasar data points satisfy the Gamma distribution in the $z$ space~\citep{Lusso2020}, we use this  Gamma distribution to generate the redshift distribution of the simulated data. At redshift $z$, the value of $\log(F_{UV})$ is assumed to satisfy the Gaussian distribution shown in Eq.~(\ref{GP}) with the mean value and the standard deviation being $ -27.503$  and $0.488$, respectively, which are obtained from the real data. We further assume that the error ($\sigma_{{UV}}$) of  $ \log (F_{UV})$  satisfies the lognormal distribution and  find,  by using the real data,  the mean value and the standard deviation being  $-4.342$  and $0.566$ respectively.
Since the real data show  a correlation between $z$ and $\log(F_{UV})$ with the  correlation coefficient being $\rho =-0.255$, we also consider this correlation  when generating the  ($z, \log(F_{UV})$) pair.
 	Then, we remove the simulated data below the UV flux limit, i.e., the minimum observable flux $\log(F_{UV})_{min}=-28.759$. The flux limit reflects the detection capability of the instrument, and quasars are unobservable by current detectors if  their  $\log(F_{UV})$ are less than $\log({F_{UV}})_{min}$.
	For any pair of ($z, \log(F_{UV})$), we generate  $\log(F_X)$  by assuming its distribution function  to be $f(\log(F_{X})) = \frac{1}{\sqrt{2 \pi}\delta}		\exp\left \{-
	\frac{[\log(F_X)-\Phi(\log(F_{UV}),d_L({\bf p}))]^2}{2\delta^2 }  \right\}$ and the luminosity relation to be the standard $L_X$-$L_{UV}$ relation, where the intrinsic dispersion $\delta$, the relation coefficients  and the PAge parameters ${\bf p}$ are taken as  fixed values shown in the third column of \autoref{Tab2}. 
	
\begin{figure}[h]
	\centering
	\includegraphics[width=.98\textwidth]{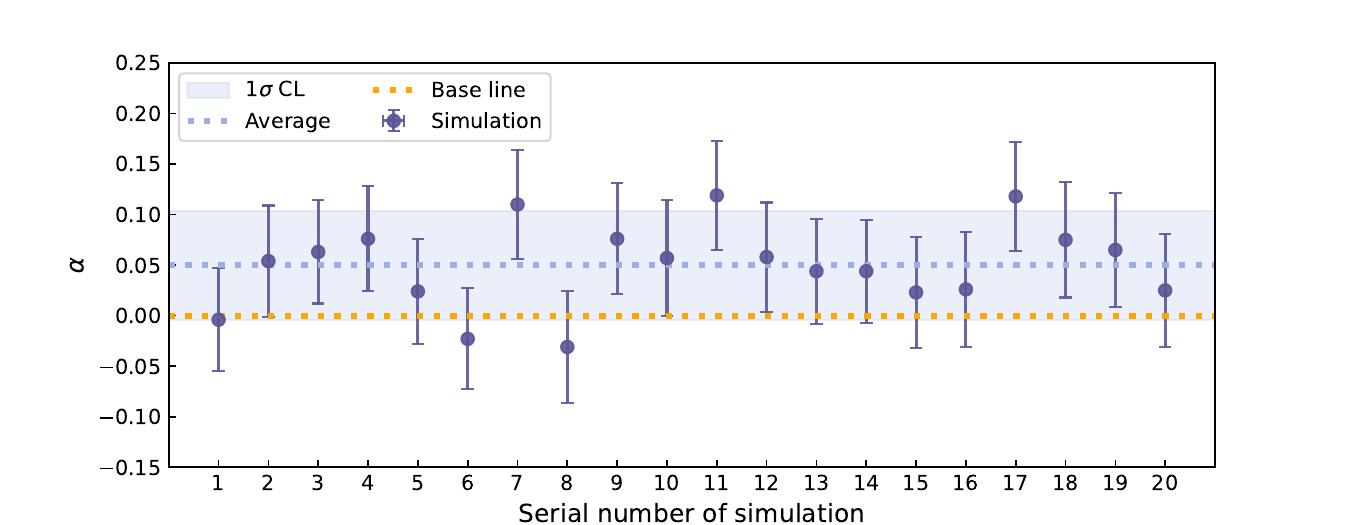}
	\caption{ The twenty results of relation coefficient $\alpha$ from the simulated data.
		The orange base line shows the $\alpha=0$ line.
		The blue dotted line represents the best fitting value of the average value $\bar{\alpha}$ and the blue shaded area corresponds to  the $1 \sigma$ CL of $\bar{\alpha}$.
			\label{fig4}
	}
\end{figure}

	With the above approach, we generate 2500 quasars. Using these simulated data, we can fit the intrinsic dispersion and the Type I relation coefficients  with the fixed PAge parameters. This process is repeated twenty times, resulting in the average values of the intrinsic dispersion and the relation coefficients to be: $\bar{\delta} = 0.2254 \pm 0.0032$, $\bar{\beta} = 6.46 \pm 0.21$, $\bar{\gamma} = 0.657 \pm 0.009$, and $\bar{\alpha} = 0.050 \pm 0.054$. 
Apparently, they have slightly smaller errors than those shown in the first column of \autoref{Tab2}. This is because  the errors of the PAge parameters are ignored when fitting the intrinsic dispersion and the relation coefficients from the simulated data.  
In \autoref{fig4},  the twenty values of the relation coefficient $\alpha$ from the simulated data are displayed to compare with  $\alpha=0$ used in the simulation.
It is easy to see that $\alpha=0$ is included in the 1$\sigma$ CL of the average value of twenty  $\alpha$. Comparing $\bar{\alpha}=0.050 \pm 0.054$ with $\alpha=0.672^{+0.063}_{
-0.073}$ shown in \autoref{Tab2}, we find that the selection effects  are ruled out as the possible cause of  redshift-evolution  of the $L_X$-$L_{UV}$ relation.

%-------------------------------------------
\section{Conclusions}\label{sec:6}

Recently, a  three-dimensional and redshift-evolutionary $L_X$-$L_{UV}$ relation of quasars was constructed from the Gaussian copula~\citep{Wang2022}.
In this paper, we compare,  by using the observational data from  quasars, the Hubble parameter measurements and the Pantheon+ SNe Ia sample,  this redshift-evolutionary  $L_X$-$L_{UV}$ relation with another redshift-evolutionary $L_X$-$L_{UV}$ relation constructed by considering a redshift dependent correction to the luminosities of quasars and the standard  $L_X$-$L_{UV}$ relation. We do not assume any cosmological model but use the PAge approximation to describe the cosmic background evolution.  The quasar data are divided into the low-redshift part and the high-redshift one. The low-redshift  and high-redshift data are then utilized respectively to constrain the relation coefficients, the SNe Ia absolute magnitude, the intrinsic dispersion of quasars and  the PAge parameters.  For the two redshift-evolutionary relations,  we find that   the low-redshift and high-redshift data can give consistent results on the PAge parameters and the absolute magnitude of the SNe Ia.   While,  the values of the relation coefficients respectively from the low-redshift and high-redshift data are discrepant.  If the standard relation is considered, all the results from the low-redshift and high-redshift data are incomparable. Furthermore, we find that the observations favor the redshift-evolutionary relations at more than $2\sigma$. Thus, the redshift-evolutionary relations of quasars are favored by the observations as opposed to the standard one.

To further compare three different relations, we use the quasar, $H(z)$ and SNe Ia data to simultaneously constrain the  PAge parameters, the intrinsic dispersion, the relation coefficients and the absolute magnitude of the SNe Ia. 
We find that  the constraints obtained with  the two redshift-evolutionary relations on the PAge parameters, the absolute magnitude, the intrinsic dispersion and the relation coefficient $\gamma$ are well consistent with each other, but they deviate significantly from those obtained in the standard relation. For the relation coefficient $\beta$, different relations lead to different values and the deviations between them are more than $2\sigma$. The observations favor strongly the redshift-evolutionary relations since  $\alpha=0$ is ruled out more than  $3\sigma$.   According to the AIC and BIC, we find that  there is strong evidence against the standard relation and mild evidence against the redshift-evolutionary relation obtained by considering a correction to luminosities of quasars.  We also confirm that  the redshift-evolution  of the  $L_X$-$L_{UV}$ relation can not be caused by the selection effects.
Therefore, we can conclude that among the three different $L_X$-$L_{UV}$ relations the redshift-evolutionary relation from copula is the most favored one.

%-------------------------------------------

\acknowledgments
We appreciate very much the insightful comments and helpful suggestions by the anonymous referee.
Bao Wang thanks Zhuoyang Li for the helpful discussions about the PAge approximation.
This work was supported in part by the NSFC under Grants No. 12275080 and No. 12075084.

\appendix

\section{PAge Approximation}\label{Appendix:PAge}

When constructing the PAge approximation,  two underlying assumptions are used~\citep{Huang2020}:
\begin{itemize}
	\item The product of the Hubble expansion rate $H(t)$ and the cosmic time $t$ can be approximated as a quadratic function of $t$.
	\item The universe is matter-dominated at redshift $z\gg 1$ with the radiation-dominant epoch, which occurs shortly before the matter domination, being disregarded.
\end{itemize}

Thus, expanding $H(t) t$ at   the cosmic age $t_0$ to the second order, we can obtain 
\begin{eqnarray}\label{ht expanse}
H(t)t&=&p_{\rm age}- (1-p_{\rm age}-p_{\rm age} q_0)(t_0-t)H_0 \nonumber \\
&-& \left(1+q_0-\frac{p_{\rm age}}{2}(2+3q_0+j_0) \right)(t_0-t)^2 H_0^2 +\mathcal{O} \left( (t_0-t)^3 \right),
\end{eqnarray}
where $p_{\rm age}\equiv H_0 t_0$, $q_0$ is the present deceleration parameter,  $j_0$ is the present jerk parameter, and  
\begin{equation}\label{factor expanse}
	a(t)=1-H_0 (t_0-t)-\frac{1}{2}H_0^2 q_0 (t_0-t)^2- \frac{1}{6}H_0^3 j_0 (t_0-t)^3+ \mathcal{O} \left( (t_0-t)^4 \right)
\end{equation}
has been used.

In order to achieve a highly precise approximation at high redshift, we now consider the second assumption.
When the redshift becomes very high ($z\gg1$) as the time tends to zero ($t\to 0$), the universe is dominated by matter.  Thus, the Hubble parameter can be expressed as $H(t)=\frac{2}{3t}$ at $t\to 0$, which means   $\lim\limits_{t\to 0}H(t)t=\frac{2}{3}$. To ensure that  Eq.~(\ref{ht expanse}) satisfies the condition  $\lim\limits_{t\to 0}H(t)t=\frac{2}{3}$, we modify the coefficient of the $(t_0-t)^2$  term and then obtain
\begin{eqnarray}
H(t)t&=&p_{age}- (1-p_{\rm age}-p_{\rm age} q_0)(t_0-t)H_0  \nonumber \\
&&- \left(1+q_0-\frac{2}{3 p_{\rm age}^2} \right)(t_0-t)^2 H_0^2.
\end{eqnarray}
Performing a mathematical transformation, one has 
\begin{equation}
	\frac{H(t)}{H_{0}}=1+\frac{2}{3}\left(1-\eta \frac{H_{0} t}{p_{\rm age}}\right)
	\left(\frac{1}{H_{0} t}-\frac{1}{p_{\rm age}}\right), 
\end{equation}
which is the common form of the PAge approximation.  Here  $\eta
\equiv 1-\frac{3}{2} p_{\rm age}^2 (1+q_0)$.

\bibliographystyle{aasjournal}
%\bibliography{reference}{}

\end{document}